# Extra Time Dimension: Deriving Five-Dimensional Relativistic Space-Time Transformations, Kinematics, and Time-Dependent Non-Relativistic Quantum Mechanics with Compactification©

Sajjad Zahir *

Professor Emeritus
University of Lethbridge, Lethbridge, AB T1K 3M4, Canada
Correspondence: Sajjad Zahir, zahir@uleth.ca, +1-403-849-2296

**Abstract**

We consider a two-time (characterized by distinct speeds of causality) and three space dimensional Minkowski space and derive relativistic coordinate and velocity transformation formulas and expressions for a new effective speed limit. Extending the ideas of Einstein's Theory of Special Relativity, concepts of five-velocity and five-momenta are introduced leading to a new formula for the rest energy of a massive object. Based on a non-relativistic limit, a two-time dependent Schrödinger-like equation is developed. Its solution for infinite square-well potential, after compactifying the extra time dimension in a closed loop topology with a period matching the Planck time, generates interference of additional quantum states with an ultra-small period of oscillation, as well. Some cosmological implications of the concept of four-dimensional versus five-dimensional masses are briefly discussed, too.

*See the last page for a brief biography and declaration of the author

## I. Introduction

The structure of space-time based on Einstein's Theory of Special Relativity (TSR) [1] and the associated four-dimensional concept of Minkowski's space [2] are familiar and well-understood. The underlying framework has one time dimension and three space dimensions (i.e., 1T + 3S space-time structure). Later, Einstein's Theory of General Relativity (TGR) introduced [3] the concept of curved space-time and a new interpretation for the gravitational force – still within the (1T + 3S) four-dimensional framework of space-time. In this paper, we will not consider gravity and thus curved space-time will remain beyond our discussion (i.e., we will consider only a 'flat' space-time). One of the earliest attempts to introduce extra dimensions (i.e., beyond the 1T + 3S four-dimensional space-time) in physics was the Kaluza-Klein Theory (KKT) that considered a five-dimensional (1T + 4S) space-time with an extra space dimension [4, 5]. But the time itself remained one-dimensional. The KKT would stand out for exploring two key ideas – 1) the unifying idea of the TGR with electro-magnetism and 2) the concept of dimensional compactification (i.e., compactifying the extra space dimension by curling it up into an ultrasmall circular loop). The current highly active field of the String Theory (ST) extended KKT ideas into many more dimensions to unify all fundamental forces including gravity (see [6] and references therein). In this paper, we will not discuss such theories. We would rather consider only one extra time dimension. Such a five-dimensional space-time with two times has been considered in the literature by several researchers from different perspectives embracing various conceptual domains (for example, see Bars [7] and references therein; also [8-12]). While considering extra time dimensions some authors think that such theories are a live conceptual possibility. If nothing else, they serve to stretch our minds into domains of new physical possibilities (see Weinstein [13, 14]).

Our focus in this paper is to explore Lorentz-like space-time transformations (see [15] and references therein) in a five-dimensional space-time with two time and three space (i.e., 2T + 3S) coordinates and the associated kinematics. Consequently, we present an intriguing conceptual framework for understanding the impact of four-dimensional versus five-dimensional masses on physical theories. Detailed work on a five-dimensional (2T + 3S) Lorentz-like transformation is not available in the literature. Our goal is to provide sufficient details so that researchers, especially the fresh entrants in the field and graduate students, can explore the formulations and compare them readily with the familiar four-dimensional Lorentz transformation results [16]. We derive the non-relativistic version matching the Newtonian mechanics. Based on the non-relativistic form and using the familiar operator formulations of quantum mechanics, a two-time dependent Schrödinger-like equation is proposed. Solving a new version of the familiar infinite square-well potential and compactifying the extra time dimension on a periodic topology we get additional secondary quantum levels. A similar example was discussed by others [6] for an extra space dimension and a Kaluza-Klein type [5] compactification technique generated additional quantum levels as well. We present an elaborate analysis and highlight the difference in results in our case.

Conceptual arguments are presented in Section II explaining the meaning of speed of causality considering extra time dimensions. In Section III, we consider a flat space-time with 2T + 3S dimensions and derive the formulas for coordinate transformation between two inertial reference frames in the standard configuration (i.e., uniform velocity is along the x-axis) with all other coordinate axes being parallel to each other. In addition, we obtain formulas for velocity transformations. Expressions for five-velocity, five-momenta, and energy-momentum relationship – all as an extension of the familiar four-dimensional (i.e., 1T + 3S space) Lorentz transformation are derived in Section IV. In Section V, we derive a non-relativistic approximation of the five-dimensional kinematics and obtain a Schrödinger-like equation in 2T + 1S dimensions for the time-dependent one-dimensional infinite square-well potential. Solutions and analysis of the quantum mechanical formulations are presented in Section VI. Comments and discussions of some unique outcomes of this research are in Section VII. Finally, Section VIII is dedicated to conclusions.

## II. The Case for Two Time Dimensions with Different Speeds of Causality

In the 1T + 3S dimension, the relativistic coordinate transformations, as we get from Einstein's TSR, have been obtained in the literature from a set of postulates [16]. In some of these sets of postulates, the constancy of the speed of light was not required (see [17] and references therein) but in the end, they needed a special speed, say *V*, that required to be 1) the maximum speed possible, 2) is tied to a particle whose rest mass is zero and, 3) is needed to be the same in all inertial frames. From empirical considerations, *V* eventually becomes the same as the speed *c* of light (i.e., photons having zero rest mass). Speed of light *c* enters the Lorentz transformation formulas through a ratio $\beta_0 = v/c$ and Lorentz factor $\gamma = 1/\sqrt{(1-\beta_0{}^2)}$ (here, *v* is the uniform speed of the inertial reference frame). As nothing can move faster than *c*, it has implications for the cause-and-effect relationship in all modes of interactions. Thus, *c* in TSR is not just the speed of light; it can be termed the speed of causality [18].

Velev [19] explores the formulations of coordinate and kinematic transformations in a detailed analysis of a flat space-time with extra time dimensions. However, Velev [19] used the same speed of causality *c* for all time dimensions including the extra ones. This at least creates one calculational problem – we cannot take limit for the extra time's *c* to zero to get back the familiar four-dimensional (i.e., those of Einstein's TSR results) formulations as there is no way to distinguish one *c* from the other. At the current energy level of the universe, we do not see the extra time dimension in any experiment. Therefore, if we think of any theory of extra time dimension it has to be about the structure of space-time at an early stage of the universe and that extra time dimension is compactified, so we do not see it now. Thus, we can conceptualize each of the time dimensions to be due to different interactions mediated by distinct massless particles moving with distinct speeds in the expanded space-time.

In such an environment, let us try to formulate the “modified” TSR for space-time through the following “gedanken” scenarios. Let us denote the interactions as interactions A and B.

First, we turn off interaction A and consider a space-time structure with 1T + 3S dimensions where interaction is carried by a massless particle moving with speed $c_1$ which is the speed of causality. Therefore, the space-time transformations (think TSR) between inertial reference frames will be the Lorentz transformations [16] with $\beta_0 = v/c_1$. Then, we turn off interaction A and turn on interaction B which will be mediated by a massless particle moving with speed $c_2$ that will be the new speed of causality (it does not have to be equal to $c_1$ because it does not know interaction A as we turned it off). Therefore, now the 1T + 3S dimensional space-time transformations between inertial reference frames will be Lorentz transformations with $\beta_0 = v/c_2$. Finally, we turn on both interactions A and B carried by respective massless particles. Now, the plausible space-time structure has to be 2T + 3S dimensional where time $t_1$ will be “influenced” by the speed of causality $c_1$ and time $t_2$ will be “influenced” by the speed of causality $c_2$.

The five-dimensional flat space-time will comprise two time dimensions each having, in general, distinct speeds of causality and three space dimensions. This is the conceptual foundation of the extra time dimensional world being considered in this paper. In the standard model of elementary particles (see Salam [20] p. 45 - 50 for a lucid explanation) there was a phase in the early universe when electromagnetic and weak interactions were carried by massless photons and W/Z bosons. However, the theory of electroweak unification is formulated in 1T + 3S dimensions, and at high energies when the gauge symmetry was unbroken massless photons and weak bosons cannot have different velocities. Therefore, we will refrain from identifying $c_1$ and $c_2$ with the speeds of massless photons and any other known interaction at all. They are just two different speeds of causality associated with time dimensions $t_1$ and $t_2$ respectively. To be more specific, we call them “speeds of causality in isolation” implying that $c_1$ is the speed of causality associated with an interaction mediated by a massless particle in the absence of any other "similar" interaction. Likewise, $c_2$ is the speed of causality associated with an interaction mediated by a massless particle in the absence of any other "similar" interaction. When both are present then we conceptualize an expanded space-time with two time dimensions in a Minkowski-like formulation having two distinct speeds of causality plus a three-dimensional space. We show in this paper that the resulting relativistic formulations involve an effective speed of causality $c_e$ that is a combination of both $c_1$ and $c_2$. It will be shown that it would be possible to travel at speeds greater than $c_1$ or $c_2$, but not greater than $c_e$.

Velev [19] derived the space-time transformation formulations and considered associated kinematics in (2T + 3S) space-time assuming the same speed of causality (i.e., the same numerical value) for both time dimensions. Consequently, Velev [19] used five-dimensional variables ($ct_1, ct_2, x, y, z$) and invariant intervals $ds^2 = c^2dt_1^2 + c^2dt_2^2 - dx^2 - dy^2 - dz^2$. He provided detailed results related to coordinate transformations, velocity, energy-momentum transformations, etc. between two inertial frames of reference. It is a natural extension of the familiar Lorentz transformation of four-dimensional Minkowski space exemplifying Einstein's TSR. He examined the causal structure of space-time showing that particles moving in multidimensional time are as stable as particles moving in one-dimensional time if certain conditions are met. Here in this paper, we consider a more general scenario – a five-dimensional space-time (i.e., two time and three space dimensions) but each time dimension has a unique speed of causality not necessarily having the same numeral value. More specifically, we make a more general assumption such that the speed of causality for time $t_1$ and $t_2$ are different. Thus, the space-time variables in the (2T + 3S) space are ($c_1t_1, c_2t_2, x, y, z$). Therefore, we can derive Velev's results [19] by taking $c_1 = c_2$ making our results a more general one. We consider Minkowski-like flat space with metric signatures (+, +, -, -, -) such that the invariant space-time interval *ds* is given by $ds^2 = c_1^2dt_1^2 + c_2^2dt_2^2 - dx^2 - dy^2 - dz^2$. With these two different speeds of causality, $c_1$ and $c_2$ we emphasize that although $t_1$ and $t_2$ are both "time-like" variables, this time ($t_1$) is not the same as that time ($t_2$). We denote the dimensions of $t_1$ and $t_2$ by T and t respectively and the dimension of *x, y, and z* by L. So, the dimensions of $c_1$ and $c_2$ are $LT^{-1}$ and $Lt^{-1}$ respectively. Throughout the paper, we keep explicitly the variables $c_1$ and $c_2$ in all expressions without assigning a value of unity (i.e., will not use *the natural system of units*). This makes it possible to set $c_2 = 0$ to obtain the four-dimensional results (i.e., those of Einstein's TSR) for the sake of comparison and for checking consistencies as needed.

**III. Space-Time Transformation in 2T + 3S Dimensions (T stands for time and S stands for space)**

We consider two times $t_1$ and $t_2$ to represent the first and second times respectively and the speeds of causality are $c_1$ and $c_2$ such that the five-dimensional space-time interval is invariant under transformation between two inertial reference frames K and K'. To be explicit,

$$\boxed{\begin{aligned} ds^2 &= c_1^2 dt_1^2 + c_2^2 dt_2^2 - dx^2 - dy^2 - dz^2 \\ ds'^2 &= c_1^2 dt_1'^2 + c_2^2 dt_2'^2 - dx'^2 - dy'^2 - dz'^2 \end{aligned}} \tag{1}$$

and $ds^2 = ds'^2$. The unprimed (primed) coordinates are defined in K (K'). We consider the standard configuration and the motion of K' is along x coordinate only such that at $t_1 = 0$ and $t_2 = 0$, the coordinate axes of K and K' coincide. K' is moving with uniform velocities $v$ and $w$ defined with respect to times $t_1$ and $t_2$ respectively. So, if $x_0$ is the coordinate of the origin of K' at any time,

$$v = dx_0 / dt_1 \; ; w = dx_0 / dt_2 \tag{2}$$

Let us assume that at times $t_1$ and $t_2$ a particle has space coordinates ($x, y, z$) in K and ($x', y', z'$) in K' corresponding to times $t_1'$ and $t_2'$. Next, we define $x_1 = ic_1t_1$, $x_2 = ic_2t_2$, $x_3 = x$, $x_4 = y$, $x_5 = z$ and $x'_1 = ic_1t_1'$, $x'_2 = ic_2t_2'$, $x'_3 = x'$, $x'_4 = y'$, $x'_5 = z'$.

To derive the transformations between K and K' we follow Velev [19]. Schröder [16] also used this technique to derive the Lorentz transformations in a more general case when the so-called Lorentz boost is in an arbitrary direction (i.e., not necessarily along x) with x, y, and z axes being parallel and coinciding space-time origins. The complete coordinate transformation will be realized by three successive rotations denoted by five-dimensional rotation matrices R, L, and $R^{-1}$. First, **R** represents a proper rotation in the $x_1$-$x_2$ plane through angle $\alpha$ giving new axes $x_{1R}$ and $x_{2R}$ keeping the other three dimensions ($x_3$, $x_4$, $x_5$) unchanged. This transformation is described by the matrix **R**,

$$\mathbf{R} = \begin{bmatrix} \cos\alpha & \sin\alpha & 0 & 0 & 0 \\ -\sin\alpha & \cos\alpha & 0 & 0 & 0 \\ 0 & 0 & 1 & 0 & 0 \\ 0 & 0 & 0 & 1 & 0 \\ 0 & 0 & 0 & 0 & 1 \end{bmatrix} \qquad (3)$$

$$x_{2R} = -x_1\sin\alpha + x_2\cos\alpha. \qquad (4)$$

Setting $x_{2R} = 0$ we get $\tan\alpha = x_2/x_1$ and so, $x_2 = x_1\tan\alpha$.

Also, rotation matrix **R** gives $x_{1R} = x_1\cos\alpha + x_2\sin\alpha$. Combining with the preceding equations, we get

$x_{1R} = x_1/\cos\alpha$

Since the reference frames K and K' are in standard configuration, at $t_1 = 0$ and $t_2 = 0$, the axes coincide. Then, to apply the boost along $x_3$ (i.e., uniform motion along $x_3$ with velocities $v$ or $w$ as defined above) a proper rotation through an angle $\phi$ in the plane $x_{1R}$-$x_3$ keeping other three dimensions ($x_{2R}$, $x_4$, $x_5$) unchanged is realized by the transformation matrix L

$$\mathbf{L} = \begin{bmatrix} \cos\phi & 0 & \sin\phi & 0 & 0 \\ 0 & 1 & 0 & 0 & 0 \\ -\sin\phi & 0 & \cos\phi & 0 & 0 \\ 0 & 0 & 0 & 1 & 0 \\ 0 & 0 & 0 & 0 & 1 \end{bmatrix} \qquad (5)$$

The origin of K' as measured in K along the $x_3$ has a value, $x_3 = v\, t_1$, or $x_3 = w\, t_2$.

Since $x_1 = ic_1t_1$ and $x_2 = ic_2t_2$ , we get, $x_1 = if_1x_3$ and $x_2 = if_2x_3$ where $f_1 = c_1/v$ and $f_2 = c_2/w$.

These relations lead to $\tan\alpha = x_2/x_1 = f_2/f_1$. Then, $\sin\alpha = f_2\beta$ and $\cos\alpha = f_2\beta$ where $\beta = \frac{1}{\sqrt{f_1^2 + f_2^2}}$.

Operating L on $x_{1R}$-$x_3$ plane, we get new coordinate value,

$$x_1' = x_{1R} \cos\phi + x_3 \sin\phi \quad (6)$$

$$x'_3 = - x_{1R} \sin\phi + x_3 \cos\phi. \quad (7)$$

Applying to the origin of K' as observed on K' (i.e., setting $x'_3 = 0$), we get $\tan\phi = x_3/ x_{1R}$. As $x_{1R} = x_1/\cos\alpha$, $x_1 = if_1x_3$ and $x_1 = if_1x_3$, we get $\tan\phi = -i\beta$, $\sin\phi = -\beta\zeta$, $\cos\phi = \zeta$, and $\zeta = \frac{1}{\sqrt{1-\beta^2}}$.

The complete transformation between K and K' is obtained by **$\Lambda$ = R$^{-1}$LR** such that **X' = $\Lambda$ X**. **X** is a column matrix with elements $[x_i]$, i=1,2,3,4,5 and **X'** is a column matrix with elements $[x'_i]$, i=1,2,3,4,5.

$$\mathbf{R}^{-1} = \begin{bmatrix} \cos\alpha & -\sin\alpha & 0 & 0 & 0 \\ \sin\alpha & \cos\alpha & 0 & 0 & 0 \\ 0 & 0 & 1 & 0 & 0 \\ 0 & 0 & 0 & 1 & 0 \\ 0 & 0 & 0 & 0 & 1 \end{bmatrix} \quad (8)$$

$$\Lambda = R^{-1}\, L\, R = \begin{bmatrix} 1+(\cos\phi-1)\cos^2\alpha & (\cos\phi-1)\sin\alpha\cos\alpha & \cos\alpha\sin\phi & 0 & 0 \\ (\cos\phi-1)\sin\alpha\cos\alpha & 1+(\cos\phi-1)\sin^2\alpha & \sin\alpha\sin\phi & 0 & 0 \\ -\cos\alpha\sin\phi & -\sin\alpha\sin\phi & \cos\phi & 0 & 0 \\ 0 & 0 & 0 & 1 & 0 \\ 0 & 0 & 0 & 0 & 1 \end{bmatrix} \quad (9)$$

Let us express all trigonometric functions in terms of $f_1$, $f_2$, $\beta$ and $\zeta$, and restore coordinate variables ($t_1$, $t_2$, $x$, $y$, $z$) in K and ($t_1'$, $t_2'$, $x'$, $y'$, $z'$) in K'. Then, we obtain the final transformation matrix as

$$\begin{bmatrix} t_1' \\ t_2' \\ x' \\ y' \\ z' \end{bmatrix} = \begin{bmatrix} A & B & C & 0 & 0 \\ D & E & F & 0 & 0 \\ G & H & I & 0 & 0 \\ 0 & 0 & 0 & 1 & 0 \\ 0 & 0 & 0 & 0 & 1 \end{bmatrix} \begin{bmatrix} t_1 \\ t_2 \\ x \\ y \\ z \end{bmatrix} \quad (10)$$

$$A = 1+\beta^2 f_1^2(\zeta-1) \qquad B = (\zeta-1) f_1 f_2 \rho \beta^2 \qquad C = -\frac{f_1}{c_1}\beta^2\zeta$$

$$D = (\zeta-1) f_1 f_2 \beta^2 \frac{1}{\rho} \qquad E = 1+\beta^2 f_2^{\,2}(\zeta-1) \qquad F = -\frac{f_2}{c_2}\beta^2\zeta$$

$$G = -c_1 f_1 \beta^2 \zeta \qquad H = -c_2 f_2 \beta^2 \zeta \qquad I = \zeta \tag{11}$$

Finally, the coordinate transformations in our five-dimensional Lorentz-like transformations are,

$$\begin{aligned} t_1' &= At_1 + Bt_2 + Cx \\ t_2' &= Dt_1 + Et_2 + Fx \\ x' &= Gt_1 + Ht_2 + Ix \\ y' &= y \\ z' &= z \end{aligned} \tag{12}$$

Here, $\rho = c_2/c_1$

That the above transformation we just derived is correct can be demonstrated by showing that

$$s'^2 = c_1^{\,2}t_1'^2 + c_2^{\,2}t_2'^2 - x'^2 - y'^2 - z'^2 = c_1^{\,2}t_1^{\,2} + c_2^{\,2}t_2^{\,2} - x^2 - y^2 - z^2 = s^2$$

We used Maplesoft™ [21] to do the algebraic computation and the outputs are given in Appendix A. It is worth deriving simplified expressions for the coordinate transformations for comparing with the familiar Lorentz transformation formulas in 1T + 3S dimension. From eq. (2) we also have

$w = \frac{dx_0}{dt_2} = \frac{dx_0}{dt_1}.\frac{dt_1}{dt_2} = v.k$, where $k = \frac{dt_1}{dt_2}$. This implies that $t_1 = k.t_2$. If $t_1$ has a dimension of T and $t_2$ has a dimension of t, then k has a dimension of $T.t^{-1}$. We will see how $k$ would help us check dimensional consistencies throughout the text of this paper. The expressions in eq. (12) simplify a lot if we use the relations, $w = v.k$ and $t_1 = k.t_2$ and we get (see Appendix B),

$$x' = \frac{x - vt_1}{\sqrt{1 - \frac{v^2}{c_e^2}}} \quad ; \quad t_1' = \frac{t_1 - \frac{xv}{c_e^2}}{\sqrt{1 - \frac{v^2}{c_e^2}}} \quad ; t_2' = \frac{t_2 - \frac{xw}{k^2 c_e^2}}{\sqrt{1 - \frac{w^2}{k^2 c_e^2}}} = \frac{t_2 - \frac{xv}{kc_e^2}}{\sqrt{1 - \frac{v^2}{c_e^2}}}$$

$$c_e^2 = c_1^2 + c_2^2 / k^2 \tag{13}$$

$$c_e^2 = c_1^2 + c_2^2 / k^2 \tag{14}$$

Results of eq. (13) also satisfy (see Appendix C),

$$s'^2 = c_1^2 t_1'^2 + c_2^2 t_2'^2 - x'^2 - y'^2 - z'^2 = c_1^2 t_1^2 + c_2^2 t_2^2 - x^2 - y^2 - z^2 = s^2$$

For real values in coordinate transformations, we should have $v \le c_e \ (\text{or}, w \le kc_e)$ and thus the maximum possible speeds $c_e$ and $kc_e$ for $v$ and $w$ respectively. So, it is possible to exceed velocity $c_1$ or $c_2$ but not $c_e$. We have tachyons [22] only for $v > c_e \ (\text{or}, w > kc_e)$. It is worth noting that,

$$t_2 - \frac{xw}{k^2 c_e^2} = \frac{1}{k}\left(t_1 - \frac{xv}{c_e^2}\right) \tag{15}$$

So, we checked that $k = \frac{t_1'}{t_2'} = \frac{t_1}{t_2}$ . Thus, *k* is invariant under the transformation of eq. (13).

One can easily derive the length contraction and time dilation formulas.

$$l = l_0 \sqrt{1 - \frac{v^2}{c_e^2}} \, ; \; \Delta t_1 = \frac{(\Delta t_1)_0}{\sqrt{1 - \frac{v^2}{c_e^2}}} ; \; \Delta t_2 = \frac{(\Delta t_2)_0}{\sqrt{1 - \frac{w^2}{k^2 c_e^2}}} = \frac{(\Delta t_2)_0}{\sqrt{1 - \frac{v^2}{c_e^2}}} \tag{16}$$

Here, $l, l_0, \Delta t_1, (\Delta t_1)_0, \Delta t_2, \text{and} (\Delta t_2)_0$ have familiar meanings requiring no further explanations as the subject is discussed in detail in all textbooks of TSR [16]. Compared to the familiar four-dimensional

Lorentz transformations, the only difference is that $c$ is replaced by $c_e$. If we set $c_2 = 0$ and $c_1 = c$, we get the familiar Einstein's TSR formulas.

Just as in traditional Lorentz transformation in the Minkowski space, it is interesting to determine how the velocities transform from the reference frames K to K'. Since we have two different time-like dimensions denoted by variables $t_1$ and $t_2$, there are two velocities for each space dimension x, y, and z. We define them as $V_x, V_y, V_z$ and $W_x, W_y, W_z$ for velocities in reference frame K with respect to $t_1$ and $t_2$ respectively. Similar quantities are defined by primed notations for the reference frame K'.

For reference frame K,

$$V_x = \frac{dx}{dt_1}, \quad W_x = \frac{dx}{dt_2}; V_y = \frac{dy}{dt_1}, \quad W_y = \frac{dy}{dt_2}; \; V_z = \frac{dz}{dt_1}, \quad W_z = \frac{dz}{dt_2} \tag{17}$$

For reference frame K'

$$V'_x = \frac{dx'}{dt'_1}, \; W'_x = \frac{dx'}{dt'_2}; V'_y = \frac{dy'}{dt'_1}, \; W'_y = \frac{dy'}{dt'_2}; V'_z = \frac{dz'}{dt'_1}, \; W'_z = \frac{dz'}{dt'_2} \tag{18}$$

Using $\frac{dt_2}{dt_1} = \frac{dt_2}{dx}\frac{dx}{dt_1} = \frac{V_x}{W_x} = \frac{V_y}{W_y} = \frac{V_z}{W_z}$ and recalling that $\frac{dt_1}{dt_2} = k$, we get $\frac{V_x}{W_x} = \frac{V_y}{W_y} = \frac{V_z}{W_z} = 1/k$.

So, $W_x = kV_x$ and so on. $\bar{W} = k\bar{V}$ and $W = |\bar{W}| = k|\bar{V}| = kV$.

Similar relations exist for the primed velocities defined in K' reference frame.

We use the results from eq. (12) and calculate the formulas for velocity transformations.

From eq. (12), we get,

$$\begin{aligned} dt_1' &= A\, dt_1 + B\, dt_2 + C\, dx \\ dt_2' &= D\, dt_1 + E\, dt_2 + F\, dx \\ dx' &= G\, dt_1 + H\, dt_2 + I\, dx \\ dy' &= dy \\ dz' &= dz \end{aligned} \tag{19}$$

$$V_x^{'} = \frac{dx^{'}}{dt_1^{'}} = \frac{G\ dt_1\ +\ H\ dt_2 +\ I\ dx}{A\ dt_1\ +\ B\ dt_2 +\ C\ dx} = \frac{G\ +\ H\ \dfrac{dt_2}{dt_1} +\ I\ \dfrac{dx}{dt_1}}{A\ +\ B\ \dfrac{dt_2}{dt_1} +\ C\ \dfrac{dx}{dt_1}} \tag{20}$$

Using

$$\frac{dt_2}{dt_1} = \frac{dt_2}{dx}\frac{dx}{dt_1} = \frac{V_x}{W_x} = \frac{V_y}{W_y} = \frac{V_z}{W_z} \tag{21}$$

We get,

$$V_x^{'} = \frac{G\ +\ H\ \dfrac{V_x}{W_x} +\ I\ V_x}{A\ +\ B\ \dfrac{V_x}{W_x} +\ C\ V_x} \tag{22}$$

Restoring expressions for G, H, I, A, B, and C and simplifying further, eq. (22) reduces to,

$$V_x^{'} = \frac{\zeta V_x \left\{ 1 - \beta^2 (\frac{c_1 f_1}{V_x} + \frac{c_2 f_2}{W_x}) \right\}}{1 + \beta^2 \frac{V_x}{v} \left\{ (\zeta - 1)(\frac{c_1 f_1}{V_x} + \frac{c_2 f_2}{W_x}) - \zeta \right\}} \tag{23}$$

Same way, we get transformations of other velocity components as

$$W_x^{'} = \frac{\zeta W_x \left\{ 1 - \beta^2 (\frac{c_1 f_1}{V_x} + \frac{c_2 f_2}{W_x}) \right\}}{1 + \beta^2 \frac{W_x}{w} \left\{ (\zeta - 1)(\frac{c_1 f_1}{V_x} + \frac{c_2 f_2}{W_x}) - \zeta \right\}} \tag{24}$$

$$V_y^{'} = \frac{V_y}{1 + \beta^2 \frac{V_x}{v} \left\{ (\zeta - 1)(\frac{c_1 f_1}{V_x} + \frac{c_2 f_2}{W_x}) - \zeta \right\}} \tag{25}$$

$$W_y^{'} = \frac{W_y}{1+\beta^2 \frac{W_x}{w}\left\{(\zeta-1)(\frac{c_1 f_1}{V_x}+\frac{c_2 f_2}{W_x})-\zeta\right\}} \tag{26}$$

$$V_z^{'} = \frac{V_z}{1+\beta^2 \frac{V_x}{v}\left\{(\zeta-1)(\frac{c_1 f_1}{V_x}+\frac{c_2 f_2}{W_x})-\zeta\right\}} \tag{27}$$

$$W_z^{'} = \frac{W_z}{1+\beta^2 \frac{W_x}{w}\left\{(\zeta-1)(\frac{c_1 f_1}{V_x}+\frac{c_2 f_2}{W_x})-\zeta\right\}} \tag{28}$$

These expressions agree with those in Velev [19] if one uses $c_1 = c_2 = c$.

These expressions can be simplified just like the coordinate transformations of eq. (13). Alternatively, as is done in the standard text on TSR (see for example [16]), we can take derivatives on both sides of the simplified transformation equations (13) and derive the velocity transformation equations easily.

$$\boxed{\begin{array}{l} V_x' = \dfrac{V_x - v}{1-\dfrac{vV_x}{c_e^{\,2}}} \text{ and } W_x' = \dfrac{W_x - w}{1-\dfrac{wW_x}{k^2 c_e^{\,2}}} = kV_x' \,;\, V_y' = \dfrac{V_y\sqrt{(1-v^2/c_e^{\,2})}}{(1-\dfrac{vV_x}{c_e^{\,2}})} \text{ and } W_y' = kV_y' \\ V_z' = \dfrac{V_z\sqrt{(1-v^2/c_e^{\,2})}}{(1-\dfrac{vV_x}{c_e^{\,2}})} \text{ and } W_z' = kV_z' \end{array}} \tag{29}$$

We can solve for $V_x$ and $W_x$ and get,

$$V_x = \frac{V_x' + v}{1+\frac{vV_x'}{c_e^{\,2}}} \text{ and } W_x = \frac{W_x' + w}{1+\frac{wW_x'}{k^2 c_e^{\,2}}} \tag{30}$$

We can confirm by substituting $c_e$ for $V_x'$ (or $kc_e$ for $W_x'$) that it is not possible to exceed the maximum speed even by traveling in a moving reference frame as in four-dimensional TSR.

$$V_x \to \frac{c_e + v}{1+\frac{vc_e}{c_e^{\,2}}} = c_e \text{ and } W_x \to \frac{kc_e + w}{1+\frac{wkc_e}{k^2 c_e^{\,2}}} = kc_e \tag{31}$$

Again, it can be checked that by setting $c_2 = 0$ and $c_1 = c$, we get the familiar Einstein's TSR formulas.

## IV. Proper time, Five velocity, Five momenta, and Relativistic Energy – Momentum Relationship

Define five space-time variables (in 2T + 3S dimensions)

$$x_1 = c_1 t_1,\ x_2 = c_2 t_2,\ x_3 = x,\ x_4 = y,\ x_5 = z$$

And the metric $g_{\mu\nu}$ is (+1, +1, -1, -1, -1) such that the displacement $s^2 = g_{\mu\nu}\, x^\mu x^\nu$ is given by,

$$s^2 = x_1^{\,2} + x_2^{\,2} - x_3^{\,2} - x_4^{\,2} - x_5^{\,2} = \left(c_1 t_1\right)^2 + (c_2 t_2)^2 - x^2 - y^2 - z^2 \tag{32}$$

We can define two proper times $d\tau_1$ and $d\tau_2$ and other related variables ($\bar{x}$ is a three-dimensional space vector with components *x, y, z*).

$$\begin{aligned}
&ds^2 = c_1^{\,2} dt_1^{\,2} + c_2^{\,2} dt_2^{\,2} - dx^2 - dy^2 - dz^2\\
&d\tau_1^{\,2} = ds^2 / c_1^{\,2} = dt_1^{\,2}\gamma_1^{-2};\ d\tau_1 = \gamma_1^{-1} dt_1\\
&d\tau_2^{\,2} = ds^2 / c_2^{\,2} = dt_2^{\,2}\gamma_2^{-2};\ d\tau_2 = \gamma_2^{-1} dt_2\\
&\gamma_1 = \frac{1}{\sqrt{1-\beta_1^2(1-\frac{1}{\beta_2^2})}};\ \gamma_2 = \frac{1}{\sqrt{1-\beta_2^2(1-\frac{1}{\beta_1^2})}};\ \bar{v}_1 = \frac{d\bar{x}}{dt_1};\ \bar{v}_2 = \frac{d\bar{x}}{dt_2}\\
&v_1 = \left|\bar{v}_1\right|, v_2 = \left|\bar{v}_1\right| \ ;\ \beta_1^2 = \frac{v_1^2}{c_1^2},\ \beta_2^2 = \frac{v_2^2}{c_2^2};\ \rho = \frac{c_2}{c_1},\ \frac{dt_1}{dt_2} = k = \frac{v_2}{v_1}
\end{aligned} \tag{33}$$

a) Five Velocity in 2T + 3S dimension

We can define two types of five velocities ($\alpha = 1,2,3,4,5$)

$$(u_1)_\alpha = \frac{dx_\alpha}{d\tau_1};(u_2)_\alpha = \frac{dx_\alpha}{d\tau_2}\ \ (x_1 = c_1 t_1,\ x_2 = c_2 t_2,\ x_3 = x,\ x_4 = y,\ x_5 = z)$$

Explicitly, one type of quantities are,

$$(u_1)_1 = c_1\gamma_1;\ (u_1)_2 = c_2\gamma_1\frac{1}{k};\ \bar{u}_1 = \gamma_1\bar{v}_1 \tag{34}$$

The other types of quantities are,

$$(u_2)_1 = c_1 k\gamma_2\,;(u_2)_2 = c_2\gamma_2\,;\bar{u}_2 = \gamma_2 \bar{v}_2 \tag{35}$$

It can be easily proved (for each type),

$$\begin{aligned}(u_1)_\alpha (u_1)^\alpha &= (u_1)_1{}^2 + (u_1)_2{}^2 - (u_1)_3{}^2 - (u_1)_4{}^2 - (u_1)_5{}^2 = c_1^2 \\ (u_2)_\alpha (u_2)^\alpha &= (u_2)_1{}^2 + (u_2)_2{}^2 - (u_2)_3{}^2 - (u_2)_4{}^2 - (u_2)_5{}^2 = c_2^2\end{aligned} \tag{36}$$

b) Energy-Momentum Five Vectors

Just like two types of five-velocities, we have two types of energy-momentum five vectors.

(37)

$(p_1)_\alpha = m_0 (u_1)_\alpha\ (\alpha = 1,2,3,4,5)$ and $m_0$ is something like **mass in five dimension**.
More explicitly,

$$(p_1)_1 = m_0 (u_1)_1 = m_0 c_1 \gamma_1 = \frac{(E_1)_1}{c_1};$$

$(E_1)_1$ can be defined as energy of type 1 corresponding to time component 1

$$(p_1)_2 = m_0 (u_1)_2 = m_0 \frac{c_2}{k}\gamma_1 = \frac{(E_1)_2}{c_2};$$

$(E_1)_2$ can be defined as energy of type 1 corresponding to time component 2

And the space components are

$$(\bar{p}_1) = m_0 \bar{u}_1 = m_0 \gamma_1 \bar{v}_1$$

It is obvious that $(p_1)^\alpha (p_1)_\alpha = m_0^2 c_1^2$ or expanding the repeated Greek index summation convention,

$$\left(\frac{(E_1)_1}{c_1}\right)^2 + \left(\frac{(E_1)_2}{c_2}\right)^2 - m_0{}^2 \bar{u}_1{}^2 = m_0^2 c_1^2$$

And for the other type, we have,

$(p_2)_\alpha = m_0 (u_2)_\alpha (\alpha = 1,2,3,4,5)$ and $m_0$ is something like **mass in five dimension**.
More explicitly,

$$(p_2)_1 = m_0 (u_2)_1 = m_0 c_1 k \gamma_2 = \frac{(E_2)_1}{c_1};$$

$(E_2)_1$ can be defined as energy of type 2 corresponding to time component 1

$$(p_2)_2 = m_0 (u_2)_2 = m_0 c_2 \gamma_2 = \frac{(E_2)_2}{c_2};$$

$(E_2)_2$ can be defined as energy of type 2 corresponding to time component 2

$$(\bar{p}_2) = m_0 \bar{u}_2 = m_0 \gamma_2 \bar{v}_2$$

It is obvious that $(p_2)^\alpha (p_2)_\alpha = m_0^2 c_2^2$ or expanding the repeated Greek index summation convention,

$$\left(\frac{(E_2)_1}{c_1}\right)^2 + \left(\frac{(E_2)_2}{c_2}\right)^2 - m_0^2 \bar{u}_2^2 = m_0^2 c_2^2 \qquad (38)$$

In addition, we observe the following relations,

$$\frac{(E_1)_1}{(E_2)_1} = \frac{(E_1)_2}{(E_2)_2} = \frac{\gamma_1}{\gamma_2 k} = \frac{1}{\rho}\ ;\ \rho\gamma_1 = \gamma_2 k$$

Also using $\rho\gamma_1 = \gamma_2 k$ we can derive $\beta_1\gamma_1 = \beta_2\gamma_2$.

Also, $\bar{v}_2 = \frac{d\bar{x}}{dt_2} = \frac{d\bar{x}}{dt_1}\frac{dt_1}{dt_2} = \bar{v}_1 k$ and $v_2 = |\bar{v}_2| = k|\bar{v}_1| = kv_1$

We have another relation, $\rho\beta_2 = k\beta_1$.

Using the above relations, it is easy to prove that the expression

$$\left(\frac{(E_1)_1}{c_1}\right)^2 + \left(\frac{(E_1)_2}{c_2}\right)^2 - m_0^2 \bar{u}_1^2 = m_0^2 c_1^2 \text{ and } \left(\frac{(E_2)_1}{c_1}\right)^2 + \left(\frac{(E_2)_2}{c_2}\right)^2 - m_0^2 \bar{u}_2^2 = m_0^2 c_2^2$$

are equivalent (see Appendix D). (39)

## V. New Expression for Rest-mass Energy and Derivation of the Non-Relativistic Limit

From the previous section, we have (using various relations)

$$\frac{(E_1)_1}{c_1} = m_0 c_1 \gamma_1 = m_0 c_1 \frac{\beta_2}{(\beta_1^2 + \beta_2^2 - \beta_1^2 \beta_2^2)^{1/2}} \tag{40}$$

$$\frac{(E_1)_2}{c_2} = m_0 \frac{c_2}{k} \gamma_1 = m_0 c_1 \frac{\beta_1}{(\beta_1^2 + \beta_2^2 - \beta_1^2 \beta_2^2)^{1/2}} \tag{41}$$

Using various identities given above, we can simplify expressions in eq. (33) and eq. (34) as we did in the case of coordinate and velocity transformations and get,

$$\frac{(E_1)_1}{c_1} + \frac{(E_1)_2}{c_2} = m_0 c_1 \frac{(\beta_1 + \beta_2)}{(\beta_1^2 + \beta_2^2 - \beta_1^2 \beta_2^2)^{1/2}} = \frac{m_0 c_1 (1 + 2c_1 c_2 / k c_e^{\ 2})^{1/2}}{\sqrt{1 - v_1^2 / c_e^{\ 2}}} \tag{42}$$

This is an interesting result of this paper (see Section VII). The rest energy of the object is given by,

$$\left( \frac{(E_1)_1}{c_1} + \frac{(E_1)_2}{c_2} \right)_{rest} = m_0 c_1 (1 + 2c_1 c_2 / k c_e^{\ 2})^{1/2} \tag{43}$$

However, in Einstein's TSR in 1T + 3D dimensions we have,

$$\left( \frac{E}{c} \right)_{rest} = mc \text{ or, } E_{rest} = mc^2 \tag{44}$$

where *c* is the speed of light and *m* is the **four-dimensional mass**.

When $kc_1 >> c_2$ or $c_2 >> c_1 k$, we get $c_e^{\ 2} \approx c_1^{\ 2}$ or $c_e^{\ 2} \approx c_2^{\ 2}/k^2$ respectively.
In addition, we have the the approximation,

$$1 + 2c_1 c_2 / k c_e^{\ 2} \to 1 \quad (\text{for } kc_1 >> c_2 \text{ or } c_2 >> kc_1)$$

The factor $k$ is introduced for the sake of dimensional matching.
We get two cases

Case 1: $kc_1 >> c_2$ which is equivalent to $\beta_1 \ll \beta_2$

$$\frac{(E_1)_1}{c_1} + \frac{(E_1)_2}{c_2} \approx m_0 c_1 \frac{1}{(1 - v_1^2 / c_1^{\ 2})^{1/2}} \tag{45}$$

For $\beta_1 \ll 1$, we have the non-relativistic limit for a free particle,

$$\frac{(E_1)_1}{c_1} + \frac{(E_1)_2}{c_2} = m_0 c_1 (1 + \frac{1}{2} \frac{v_1^2}{c_1^{\ 2}}) \tag{46}$$

Case 2: $c_2 >> kc_1$ which is equivalent to $\beta_2 \ll \beta_1$

$$\frac{(E_1)_1}{c_1}+\frac{(E_1)_2}{c_2}\approx m_0 c_1 \frac{1}{(1-v_2^2/{c_2}^2)^{1/2}} \tag{47}$$

For $\beta_2 \ll 1$, we have the non-relativistic limit for a free particle,

$$\frac{(E_1)_1}{c_1}+\frac{(E_1)_2}{c_2}= m_0 c_1 (1+\frac{1}{2}\frac{v_2^2}{{c_2}^2}) \tag{48}$$

Similar expressions can be obtained using $(E_2)_1 \text{and} (E_2)_2$. However, for the sake of illustration, we focus on Case 1 and derive two time-dependent Schrödinger-like equations (i.e., the one-dimensional infinite square-well potential problem in the 2T+1S dimension). Eventually, the extra time dimension will be compactified. We start with the non-relativistic limit eq. (36),

$$\frac{(E_1)_1}{c_1}+\frac{(E_1)_2}{c_2}= m_0 c_1 + \frac{1}{2} m_0 c_1 \frac{v_1^2}{c_1^2} = m_0 c_1 + \frac{1}{2} m_0 \frac{v_x^2}{c_1} = \frac{1}{c_1}[(m_0 c_1^2)+\frac{(p_x)^2}{2m_0}] \tag{49}$$

$p_x$ is the non-relativistic linear momentum $m_0 v_x$. The first term in the square bracket is the rest-mass energy that we drop from now on and the second term is the kinetic energy. Adding a potential energy term *V(x)* we get,

$$\frac{(E_1)_1}{c_1}+\frac{(E_1)_2}{c_2}=\frac{1}{c_1}[\frac{(p_x)^2}{2m_0}+V(x)] \tag{50}$$

Next, by replacing the classical energy and momentum functions with corresponding quantum operators we get the time-dependent Schrödinger-like equation.

$$(E_1)_1 \rightarrow i\hbar\frac{\partial}{\partial t_1},\ (E_1)_2 \rightarrow i\hbar\frac{\partial}{\partial t_2},\ p_x \rightarrow -i\hbar\frac{\partial}{\partial x} \tag{51}$$

$$i\hbar\frac{\partial}{c_1\partial t_1}\Psi(t_1,t_2,x)+i\hbar\frac{\partial}{c_2\partial t_2}\Psi(t_1,t_2,x)=-\frac{\hbar^2}{2m_0 c_1}\frac{\partial^2}{\partial x^2}\Psi(t_1,t_2,x)+\frac{1}{c_1}V(x)\Psi(t_1,t_2,x) \tag{52}$$

## VI. Solving the Infinite Square-Well Potential Problem in 2T + 1S dimension with Compactification of the Extra Time Dimension and Analysis of Results

We consider one-dimensional infinite square-well potential,

$$V(x) = 0;\ 0 \le x \le a\ ;\ V(x) = \infty;\ 0 \ge x \ge a \tag{53}$$

Such an example is well discussed in the literature as a time-independent Schrödinger equation in 1T + 2S dimension where the extra space dimension is compactified as in Kaluza-Klein theory [4, 5] on a circle of radius R (for example, see Zwiebach [6]). The energy eigenvalues are determined by two quantum numbers *q* and *l*.

$$E_{k,l} = \frac{\hbar^2}{2m}[(\frac{q\pi}{a})^2 + (\frac{l}{R})^2]; q = 1,2,... \text{ and } l = 0,1,2,.. \tag{54}$$

The term involving *q* defines the familiar quantum effect in an infinite square-well problem, whereas the term involving *l* is due to the compactification of the extra space dimension. We will not discuss this result further as it is available in the literature. We focus on our example with an extra time dimension that will be compactified on a circle of period $T_0$. The solution will be obtained by following the standard separation of variable techniques.

$$\Psi(t_1, t_2, x) = \psi(x)\, X(t_1, t_2) \tag{55}$$

The time-dependent equation becomes,

$$\frac{1}{X(t_1,t_2)}[i\hbar\frac{\partial}{c_1\partial t_1}X(t_1,t_2) + i\hbar\frac{\partial}{c_2\partial t_2}X(t_1,t_2)] = \frac{1}{\psi(x)}[-\frac{\hbar^2}{2m_0c_1}\frac{\partial^2}{\partial x^2}\psi(x) + \frac{1}{c_1}V(x)\psi(x)] \tag{56}$$

This implies that both sides will be equal to a constant $\frac{E}{c_1}$ (this choice leads to simple equations).

$$i\hbar\frac{\partial}{c_1\partial t_1}X(t_1,t_2) + i\hbar\frac{\partial}{c_2\partial t_2}X(t_1,t_2) = \frac{E}{c_1}X(t_1,t_2) \tag{57}$$

$$-\frac{\hbar^2}{2m_0}\frac{\partial^2}{\partial x^2}\psi(x) + V(x)\psi(x) = E\psi(x) \tag{58}$$

We further apply the separation of variable technique to the eq. (57).

Let $X(t_1,t_2)$ = $T_1(t_1)T_2(t_2)$ and derive two equations

$$i\hbar\frac{\partial}{c_1\partial t_1}T_1(t_1)=\frac{\omega_1}{c_1}T_1(t_1) \tag{59}$$

$$i\hbar\frac{\partial}{c_2\partial t_2}T_2(t_2)=\frac{\omega_2}{c_2}T_2(t_2) \tag{60}$$

where,

$$\frac{\omega_1}{c_1}+\frac{\omega_2}{c_2}=\frac{E}{c_1} \tag{61}$$

Solving the equations for $T_1(t_1)$ and $T_2(t_2)$ we get,

$$T_1(t_1)=Ae^{-i\frac{\omega_1}{\hbar}t_1}\ ;\ T_2(t_2)=Be^{-i\frac{\omega_2}{\hbar}t_2}$$
$$A \text{ and } B \text{ are arbitrary constants.} \tag{62}$$

We assume that time $t_2$ is compactified in the sense that,

$T_2(t_2)=T_2(t_2+T_0)$ and $T_0$ is very small, say like Planck time. This gives,

$$e^{-i\frac{\omega_2}{\hbar}T_0}=1=e^{i2p\pi}\ ; p=0,\pm1,\pm2,..$$
$$\omega_2=-\frac{2p\pi\hbar}{T_0} \tag{63}$$

For the sake of illustration, we take the positive values. Now, let us consider the x-dimensional equation.

$$-\frac{\hbar^2}{2m_0}\frac{\partial^2}{\partial x^2}\psi(x)+V(x)\psi(x)=E\psi(x) \tag{64}$$

We introduce the variable,

$$q=\sqrt{\frac{2m_0E}{\hbar^2}} \tag{65}$$

Then, the solution that satisfies boundary conditions at *x* = 0 and *x* = a is well-known and available in any introductory quantum mechanics textbook (see for example [23]),

$$\psi(x)=C\sin(qx);\ q_n=\frac{n\pi}{a},\ n=1,2,... \tag{66}$$

Energy levels are $E_n = \frac{\hbar^2 n^2 \pi^2}{2m_0 a^2}$ (67)

Finally, using $\frac{\omega_1}{c_1} + \frac{\omega_2}{c_2} = \frac{E}{c_1}$ (i.e., eq. (61))

we get,

$$\frac{\omega_1}{c_1} = \frac{E}{c_1} - \frac{\omega_2}{c_2} = \frac{\hbar^2 n^2 \pi^2}{2m_0 a^2 c_1} + \frac{2p\pi\hbar}{T_0 c_2} \tag{68}$$

Or,

$$\frac{\omega_1(n,p)}{c_1} = \frac{\hbar^2 n^2 \pi^2}{2m_0 a^2 c_1} + \frac{2p\pi\hbar}{T_0 c_2}; n = 1, 2, ..\text{and } p = 0, \pm 1, \pm 2, .. \tag{69}$$

The complete solution is a linear combination of the products.

$$\Psi(t_1, t_2, x) = \sum_{n,p} C(n,p)\psi_n(x) T_{1p}(t_1) T_{2(n,p)}(t_2) \tag{70}$$

$C(n, p)$ are arbitrary constants. For the sake of analysis and to highlight the results, we consider a couple of cases as follows.

Case 1:

Let *p* =1 and *n* = 1, 2 and consider the linear combination of two terms only.

$$\Psi_1 = C_1 \sin q_1 x . A_{1,1} e^{-i\frac{\omega_1(1,1)}{\hbar}t_1} . B_1 e^{-i\frac{\omega_2(1)}{\hbar}t_2} = C(1,1) \sin q_1 x . e^{-i\frac{\omega_1(1,1)}{\hbar}t_1} . e^{-i\frac{\omega_2(1)}{\hbar}t_2} \tag{71}$$

Similarly,

$$\Psi_2 = C(1,2) \sin q_2 x . e^{-i\frac{\omega_1(1,2)}{\hbar}t_1} . e^{-i\frac{\omega_2(1)}{\hbar}t_2} \tag{72}$$

All constants are lumped into *C*(1,1) and *C*(1,2) and we assume them to be real for the sake of simplicity. The sum of the two solutions gives,

$$\Psi = \Psi_1 + \Psi_2 = \left( C(1,1) \sin q_1 x . e^{-i\frac{\omega_1(1,1)}{\hbar}t_1} + C(1,2) \sin q_2 x . e^{-i\frac{\omega_1(1,2)}{\hbar}t_1} \right) e^{-i\frac{\omega_2(1)}{\hbar}t_2} \tag{73}$$

Therefore,

$$
\begin{aligned}
|\Psi|^2 &= C(1,1)^2 \sin^2 q_1 x + C(1,2)^2 \sin^2 q_2 x + \\
&\quad 2C(1,1)C(1,2)\sin q_1 x . \sin q_2 x . \cos(\omega_1(1,1) - \omega_1(1,2))\frac{t_1}{\hbar} \\
&= C(1,1)^2 \sin^2 q_1 x + C(1,2)^2 \sin^2 q_2 x + 2C(1,1)C(1,2)\sin q_1 x . \sin q_2 x . \cos(\frac{3\hbar\pi^2}{2m_0 a^2})t_1
\end{aligned} \tag{74}
$$

This gives the time evolution with respect to time $t_1$. Dependence on $t_2$ did not appear due to the particular choice of the quantum numbers.

Case 2:

However, let us consider another case where *p* = 1, 2, and *n* = 1.

The two quantum states under consideration are,

$$
\Psi_1 = D_1 \sin q_1 x . e^{-i\frac{\omega_1(1,1)}{\hbar}t_1} . e^{-i\frac{\omega_2(1)}{\hbar}t_2} \tag{75}
$$

and

$$
\Psi_2 = D_2 \sin q_1 x . e^{-i\frac{\omega_1(2,1)}{\hbar}t_1} . e^{-i\frac{\omega_2(2)}{\hbar}t_2} \tag{76}
$$

All constants are lumped into $D_1$ and $D_2$ and considering them to be real for the sake of simplicity, the sum of the solutions gives,

$$
\Psi = \Psi_1 + \Psi_2 = \sin q_1 x \left( D_1 e^{-i\frac{\omega_1(1,1)}{\hbar}t_1} . e^{-i\frac{\omega_2(1)}{\hbar}t_2} + D_2\, e^{-i\frac{\omega_1(2,1)}{\hbar}t_1} . e^{-i\frac{\omega_2(2)}{\hbar}t_2} \right) \tag{77}
$$

Therefore,

$$
\begin{aligned}
|\Psi|^2 &= \sin^2 q_1 x \left\{ D_1^{\,2} + D_2^{\,2} + 2D_1 D_2 \cos[(\omega_1(1,1) - \omega_1(2,1))\frac{t_1}{\hbar} + (\omega_2(1) - \omega_2(2))\frac{t_2}{\hbar}] \right\} \\
&= \sin^2 q_1 x \left\{ D_1^{\,2} + D_2^{\,2} + 2D_1 D_2 \cos[2\pi(\frac{c_1 t_1}{T_0 c_2} + \frac{t_2}{T_0})] \right\}
\end{aligned} \tag{78}
$$

This result has dependence in both $t_1$ and $t_2$. However, we invoke the relation $t_1 = kt_2$ and define $\lambda = c_2 / (c_1 k)$ which is a dimensionless quantity.

$$\left|\Psi\right|^2 = \sin^2 q_1 x \left\{ D_1^{\,2} + D_2^{\,2} + 2D_1 D_2 \cos[2\pi(\frac{1+\lambda}{\lambda k T_0})t_1] \right\} \tag{79}$$

So, the frequency $f$ of oscillation is $\frac{1+\lambda}{\lambda k T_0}$ and the period $P$ is $\frac{\lambda k T_0}{1+\lambda}$. We assumed that $t_2$ was compactified such that $t_2 = t_2 + T_0$. We remember that $kT_0$ has the same dimension as $t_1$ and we can take its value to be equal to Planck time $T_p$ whose value is 5.39 x $10^{-44}$ sec. Thus,

$$P / T_p = \frac{\lambda}{1+\lambda} \tag{80}$$

We can plot $P / T_p$ against $\lambda$ and see how it varies as $\lambda$ (see Figure 1). When $\lambda = 0$ (i.e., $c_2 = 0$), we get zero value for a period as it reduces to four-dimensional spacetime (i.e., 1T + 3S) making compactification of the extra time dimension redundant. The smallest time measured so far is about 2.5 x $10^{-19}$ seconds [24]. We need to point out that this analysis is based on a specific example that was chosen to demonstrate the overlapping effect of the non-relativistic compactified quantum states and on the compactification method.

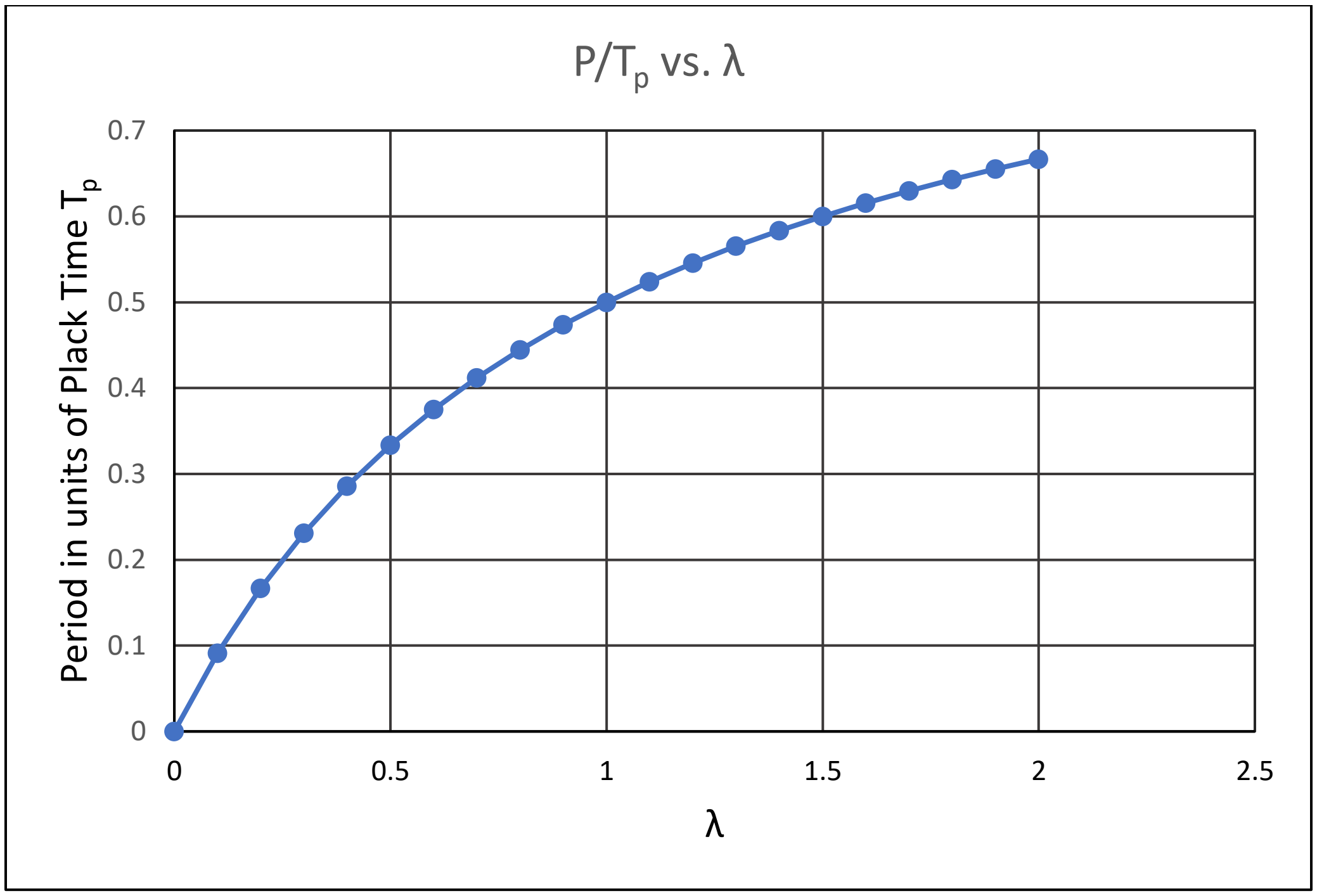

**Figure 1: Period of Oscillation in Units of Planck Time vs λ**

## VII. Comments and Discussion

We are considering relativistic transformations in a 2T + 3S space with the square of interval given by,

$$ds^2 = c_1^{\,2}dt_1^{\,2} + c_2^{\,2}dt_2^{\,2} - dx^2 - dy^2 - dz^2 \tag{81}$$

We defined $\dfrac{dt_1}{dt_2} = k$ and *k* is considered a constant. It is the same as $1/\xi$ used by Quiros [25]. The scale factor is invariant under coordinate transformations, i.e., Under this transformation, the interval can be re-expressed as,

$$ds^2 = (c_1^{\,2} + c_2^{\,2}/k^2)dt_1^{\,2} - dx^2 - dy^2 - dz^2 \; = c_e^{\,2}dt_1^{\,2} - dx^2 - dy^2 - dz^2 \tag{82}$$

whereas defined earlier. This is the same interval as in Einstein's TSR with *c* being replaced by $c_e$. Therefore, in analogy with the familiar Lorentz transformation, formulas for coordinate, velocity, momentum, etc. are expected to be identical to those in TSR except *c* being replaced by $c_e$ in our case. That's why we got similar results in the previous sections. To be more specific, all the expressions of relativistic transformation formulas reduce to the familiar Lorentz ones as $c_2 \to 0$. We can call this a *reduced or effective interval* and the scale transformation can be called a *static compactification technique* at the coordinate level. However, the energy-momentum four (or five) vectors need some comment.

The reduced four momenta will satisfy, $p_\mu p^\mu = m^2 c_e^2$ where m is like a **four-dimensional mass**.

Therefore,

$$\boxed{\begin{aligned} & p_\mu = (p_0, \bar{p}) = (E/c_e, \bar{p}) \\ & E = mc_e^{\,2}\gamma_e\,;\, \bar{p} = m\bar{v}\gamma_e\,;\gamma_e = 1/\sqrt{(1 - v^2/c_e^{\,2})} \\ & \left(\frac{E}{c_e}\right)_{rest} = mc_e \end{aligned}} \tag{83}$$

In our case, we obtained (see eq. (43)),

$$\left(\frac{(E_1)_1}{c_1}+\frac{(E_1)_2}{c_2}\right)_{rest}=m_0c_1(1+2c_1c_2/kc_e^{\,2})^{1/2} \tag{84}$$

If we assume that the right-hand sides of eq. (83) and eq. (84) are equivalent quantities, then we get,

$$\frac{m_0}{m}=\frac{c_e}{c_1(1+2c_1c_2/kc_e^{\,2})^{1/2}}=\sqrt{\frac{1+c_2^{\,2}/(c_1k)^2}{1+\dfrac{2c_2}{c_1k(1+c_2^{\,2}/(c_1k)^2)}}}=\sqrt{\frac{1+\lambda^2}{1+\dfrac{2\lambda}{(1+\lambda^2)}}} \tag{85}$$

where $\lambda = c_2/(c_1k)$ which is a dimensionless quantity. When both masses are four-dimensional and are equal.

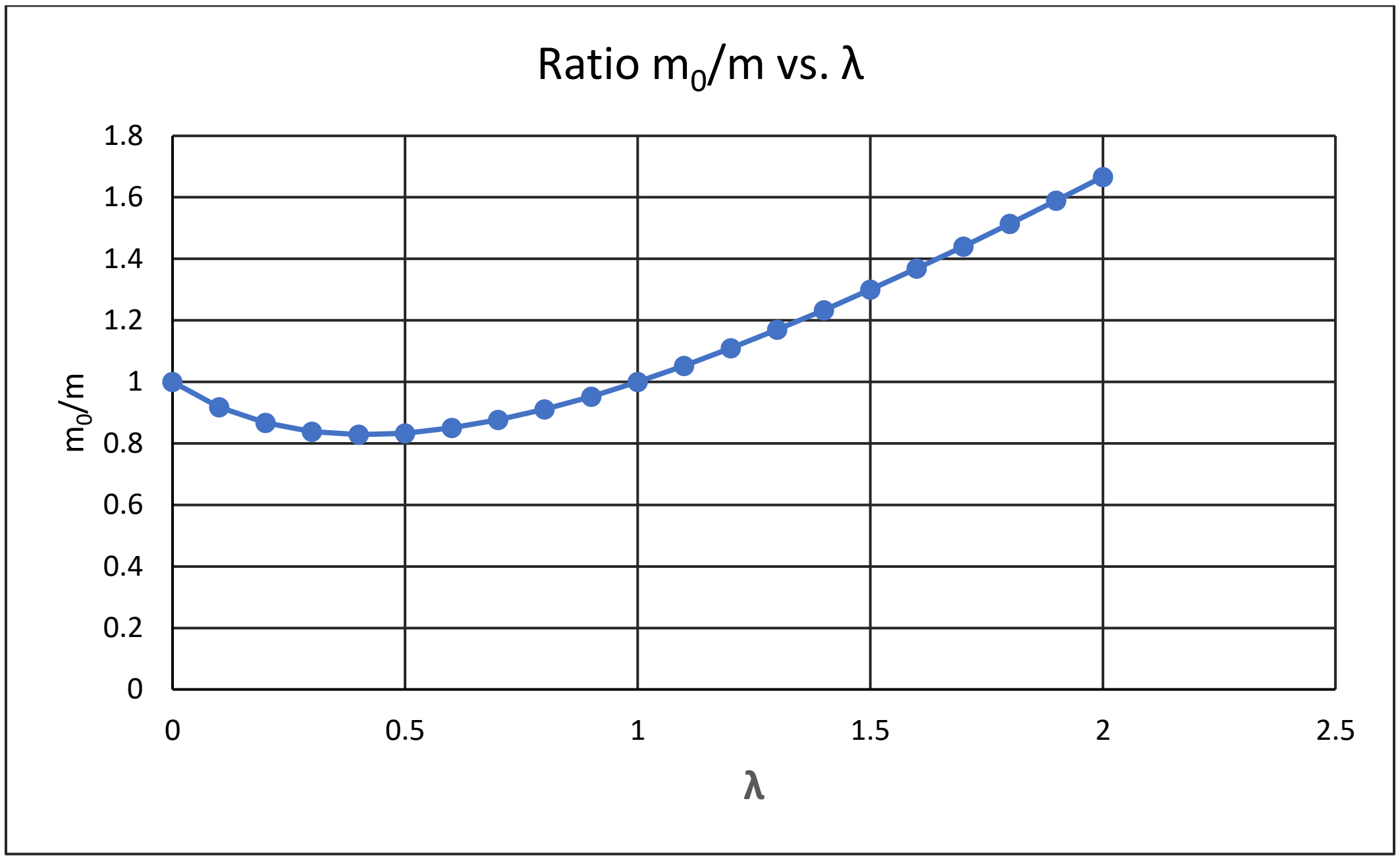

**Figure 2: Ratio of Five-dimensional Mass to Four-dimensional Mass vs. λ**

We plotted the mass ratio against λ and see the interesting behavior (Figure 2). In the range $0<\lambda<1$, the ratio $m_0/m$ is less than 1 and is greater than 1 beyond. Does it imply that even the rest mass of an object depends on the finer details of the dimensional structure of space and time?

On the other hand, if we assume that the four-dimensional and five-dimensional masses are the same, then the rest energies denoted by $(P_t)^4{}_{rest}$ and $(P_t)^5{}_{rest}$ are different. Here the subscript *t* is chosen to imply that $P_t$ stands for the "time component" of the corresponding four or five momenta. From eq. (83) and eq. (84), we get,

$$\left(P_t\right)_{rest}{}^{4} = \left(\frac{E}{c_e}\right)_{rest} = mc_e \tag{86}$$

$$(P_t)^5{}_{rest} = \left(\frac{(E_1)_1}{c_1} + \frac{(E_1)_2}{c_2}\right)_{rest} = m_0 c_1 (1 + 2c_1 c_2 / kc_e{}^2)^{1/2} \tag{87}$$

Then, we have ($m = m_0$)

$$\frac{(P_t)^4{}_{rest}}{(P_t)^5{}_{rest}} = \frac{c_e}{c_1(1+2c_1c_2/kc_e{}^2)^{1/2}} = \sqrt{\frac{1+c_2{}^2/(c_1k)^2}{1+\dfrac{2c_2}{c_1k(1+c_2{}^2/(c_1k)^2)}}} = \sqrt{\frac{1+\lambda^2}{1+\dfrac{2\lambda}{(1+\lambda^2)}}} \tag{88}$$

The plot of eq. (88) will be just as Figure 2 except that the vertical axis will imply $\frac{(P_t)^4{}_{rest}}{(P_t)^5{}_{rest}}$. If we think in terms of cosmological phenomena, how can we explain the variation of the ratio of the five-dimensional mass to four-dimensional mass or $\frac{(P_t)^4{}_{rest}}{(P_t)^5{}_{rest}}$ with respect to λ and compactification in the context of (rest mass) energy release or absorption spontaneously during the early evolution of the universe?

The *static compactification* at the coordinate level has a weakness as it will prevent further expansion of physical theories into the dynamic space (i.e., classical and quantum field theory). The 2T + ö1S dimensional square-well Schrödinger-like equation is a good example. Because of the second time dimension in the formulation of the wave equation, we could compactify the extra time dimension on a circular topology and derive additional secondary quantum energy levels. If the extra time variable is not available, we cannot develop field theoretical formulations as it will not be possible to couple with additional fifth tie components or their corresponding derivatives (e.g., five-dimensional Klein-Gordon field theory or equivalent Abelian gauge theories).

## VIII. Conclusions and Future Research

This is an expanded version of the paper [26] where we considered the space-time structure with two time and three space dimensions. Other researchers also investigated the possibility of two-time physics from different considerations. Our conceptual framework for the scientific considerations of this paper refers to the early stage of the universe when all particles of the standard model were massless. We postulated that the present era (i.e., current energy scale) 1T + 3S dimensional special theory of relativity (TSR) can be extended to a 2T+3S dimensional TSR applicable to a scenario where each time dimension was tied to a distinctly separate speed of causality $c_1$ and $c_2$. That scenario contrasts the present era one with one speed of causality $c$, the speed of light. In our conceptual framework speeds $c_1$ and $c_2$ are tied to some fundamental interactions mediated by some massless particles that carry information needed for implementing the "cause-and-effect" phenomenon.

We derived formulations for coordinate transformation, velocity transformation, and energy-momentum relations in the five-dimensional (i.e., 2T + 3S) spacetime and showed (see Appendices) the usefulness of MappleSoft ™. We found that the maximum speed possible is $c_e = \sqrt{\left(c_1^2 + c_2^2 / k^2\right)}$ (k is defined in the text) which is greater than either $c_1$ or $c_2$ without violating the "new TSR". The term $c_e$ is the speed of any particle having non-zero energy but a zero rest mass. The rest energy of any particle is given by eq. (43). This can be compared with the four-dimensional TSR expression eq. (44). Starting from the five-dimensional energy-momentum relation, we derived the "non-relativistic" limit from which the time-dependent Schrödinger-like equation (TDE) of the non-relativistic quantum mechanics is derived for a 2T + 1S dimension.

Our consideration of an extra time dimension was conceptualized at the very early stage of the creation of the universe. We anticipate this to be after the inflationary expansion and during the "re-heating" phase when all the standard particles were created, albeit in a mass-less form. However, as in the current energy range, we only see a 1T + 3S space-time configuration the four-dimensional STR works well. Thus, the extra time dimension is expected to be compactified. In the context of the Kaluza-Klein theory, compactification is done on an ultrasmall circular topology (either spatial or temporal dimension). However, we have not explored if space-time can be compactified (or expanded) in other, perhaps more dynamic, ways [27]. For example, cosmologists mention "metric expansion" to explain the inflationary phase by conceptualizing the existence of a scalar field called inflaton [28]. But how that works mathematically or in theoretical formulations is not known. The Discovery of the Higgs particle confirms theoretical developments including symmetry-breaking in electroweak unification. But we do not know how the Higgs field can play any role in compactifying the extra time dimension. Conceptually, if inflaton can be responsible for metric expansion during the inflationary phase, making the Higgs field responsible for compactifying the extra time dimension may not be too far-fetched in imagination [29].

**Appendix A**

Showing $s'^2 = c_1^2 t_1'^2 + c_2^2 t_2'^2 - x'^2 - y'^2 - z'^2 = c_1^2 t_1^2 + c_2^2 t_2^2 - x^2 - y^2 - z^2 = s^2$

using general expressions of eq. (10) and Maplesoft™ [21]. Maplesoft key symbols X, T, TT, XP, TP, and TTP represent text variables x, $t_1$, $t_2$, x', $t'_1$, $t'_2$, x' respectively.

> $f1 := \frac{c1}{v};$

$$f1 := \frac{c1}{v} \quad (1)$$

> $f2 := \frac{c2}{w};$

$$f2 := \frac{c2}{w} \quad (2)$$

> $R := \frac{c2}{c1};$

$$R := \frac{c2}{c1} \quad (3)$$

> $B := \frac{1}{\sqrt{f1^2 + f2^2}};$

$$B := \frac{1}{\sqrt{\frac{c1^2}{v^2} + \frac{c2^2}{w^2}}} \quad (4)$$

> $Z := \frac{1}{\sqrt{1 - B^2}};$

$$Z := \frac{1}{\sqrt{-\frac{1}{\frac{c1^2}{v^2} + \frac{c2^2}{w^2}} + 1}} \quad (5)$$

$$> TP := \left(1+(Z-1)\cdot B^2\cdot f1^2\right)\cdot T + R\cdot(Z-1)\cdot f1\cdot f2\cdot B^2\cdot TT + \left(-\frac{B^2\cdot Z\cdot f1}{c1}\right)\cdot X;$$

$$TP := \left(1 + \frac{\left(\frac{1}{\sqrt{-\frac{1}{\frac{c1^2}{v^2}+\frac{c2^2}{w^2}}+1}}-1\right)c1^2}{\left(\frac{c1^2}{v^2}+\frac{c2^2}{w^2}\right)v^2}\right)T + \frac{c2^2\left(\frac{1}{\sqrt{-\frac{1}{\frac{c1^2}{v^2}+\frac{c2^2}{w^2}}+1}}-1\right)TT}{v\,w\left(\frac{c1^2}{v^2}+\frac{c2^2}{w^2}\right)} - \frac{X}{\left(\frac{c1^2}{v^2}+\frac{c2^2}{w^2}\right)\sqrt{-\frac{1}{\frac{c1^2}{v^2}+\frac{c2^2}{w^2}}+1}\;v} \tag{6}$$

$$> TTP := \frac{(Z-1)}{R}\cdot f1\cdot f2\cdot B^2\cdot T + \left(1+(Z-1)\cdot f2^2\cdot B^2\right)\cdot TT + \left(-\frac{B^2\cdot Z\cdot f2}{c2}\right)\cdot X;$$

$$TTP := \frac{\left(\frac{1}{\sqrt{-\frac{1}{\frac{c1^2}{v^2}+\frac{c2^2}{w^2}}+1}}-1\right)c1^2\,T}{v\,w\left(\frac{c1^2}{v^2}+\frac{c2^2}{w^2}\right)} + \left(1 + \frac{\left(\frac{1}{\sqrt{-\frac{1}{\frac{c1^2}{v^2}+\frac{c2^2}{w^2}}+1}}-1\right)c2^2}{w^2\left(\frac{c1^2}{v^2}+\frac{c2^2}{w^2}\right)}\right)TT - \frac{X}{\left(\frac{c1^2}{v^2}+\frac{c2^2}{w^2}\right)\sqrt{-\frac{1}{\frac{c1^2}{v^2}+\frac{c2^2}{w^2}}+1}\;w} \tag{7}$$

> $XP := \left(-B^2 \cdot c1 \cdot f1 \cdot Z\right) \cdot T + \left(-B^2 \cdot f2 \cdot c2 \cdot Z\right) \cdot TT + Z \cdot X;$

$$XP := -\frac{c1^2\, T}{\left(\frac{c1^2}{v^2} + \frac{c2^2}{w^2}\right) v \sqrt{-\frac{1}{\frac{c1^2}{v^2} + \frac{c2^2}{w^2}} + 1}} - \frac{TT\, c2^2}{\left(\frac{c1^2}{v^2} + \frac{c2^2}{w^2}\right) \sqrt{-\frac{1}{\frac{c1^2}{v^2} + \frac{c2^2}{w^2}} + 1}\; w} + \frac{X}{\sqrt{-\frac{1}{\frac{c1^2}{v^2} + \frac{c2^2}{w^2}} + 1}} \quad \textbf{(8)}$$

> $INTV := (c1 \cdot TP)^2 + (c2 \cdot TTP)^2 - XP^2;$

$$
INTV := c1^2 \Bigg( \Bigg( 1 + \frac{\left( \frac{1}{\sqrt{-\frac{1}{\frac{c1^2}{v^2} + \frac{c2^2}{w^2}} + 1}} - 1 \right) c1^2}{\left( \frac{c1^2}{v^2} + \frac{c2^2}{w^2} \right) v^2} \Bigg) T + \frac{c2^2 \left( \frac{1}{\sqrt{-\frac{1}{\frac{c1^2}{v^2} + \frac{c2^2}{w^2}} + 1}} - 1 \right) TT}{v\, w \left( \frac{c1^2}{v^2} + \frac{c2^2}{w^2} \right)} - \frac{X}{\left( \frac{c1^2}{v^2} + \frac{c2^2}{w^2} \right) \sqrt{-\frac{1}{\frac{c1^2}{v^2} + \frac{c2^2}{w^2}} + 1}\; v} \Bigg)^2 + c2^2 \Bigg( \frac{\left( \frac{1}{\sqrt{-\frac{1}{\frac{c1^2}{v^2} + \frac{c2^2}{w^2}} + 1}} - 1 \right) c1^2\, T}{v\, w \left( \frac{c1^2}{v^2} + \frac{c2^2}{w^2} \right)} + \Bigg( 1 + \frac{\left( \frac{1}{\sqrt{-\frac{1}{\frac{c1^2}{v^2} + \frac{c2^2}{w^2}} + 1}} - 1 \right) c2^2}{w^2 \left( \frac{c1^2}{v^2} + \frac{c2^2}{w^2} \right)} \Bigg) TT \qquad (9)
$$

$$- \frac{X}{\left(\frac{c1^2}{v^2} + \frac{c2^2}{w^2}\right) \sqrt{-\frac{1}{\frac{c1^2}{v^2} + \frac{c2^2}{w^2}} + 1}\; w} \Bigg)^2 - \Bigg($$

$$- \frac{c1^2\, T}{\left(\frac{c1^2}{v^2} + \frac{c2^2}{w^2}\right) v \sqrt{-\frac{1}{\frac{c1^2}{v^2} + \frac{c2^2}{w^2}} + 1}}$$

$$- \frac{TT\, c2^2}{\left(\frac{c1^2}{v^2} + \frac{c2^2}{w^2}\right) \sqrt{-\frac{1}{\frac{c1^2}{v^2} + \frac{c2^2}{w^2}} + 1}\; w} + \frac{X}{\sqrt{-\frac{1}{\frac{c1^2}{v^2} + \frac{c2^2}{w^2}} + 1}} \Bigg)^2$$

> *fint* := *simplify*(*INTV*);

$$fint := T^2\, c1^2 + TT^2\, c2^2 - X^2 \qquad \textbf{(10)}$$

>

**Appendix B**

To get the simplified form eq. (13) from eq. (10), we used Maplesoft™ [21].

> $f1 := \frac{c1}{v};$

$$f1 := \frac{c1}{v} \quad (1)$$

> $f2 := \frac{c2}{w};$

$$f2 := \frac{c2}{k\,v} \quad (2)$$

> $R := \frac{c2}{c1};$

$$R := \frac{c2}{c1} \quad (3)$$

> $B := \frac{1}{\sqrt{f1^2 + f2^2}};$

$$B := \frac{1}{\sqrt{\frac{c1^2}{v^2} + \frac{c2^2}{k^2\,v^2}}} \quad (4)$$

> $Z := \frac{1}{\sqrt{1 - B^2}};$

$$Z := \frac{1}{\sqrt{-\frac{1}{\frac{c1^2}{v^2} + \frac{c2^2}{k^2\,v^2}} + 1}} \quad (5)$$

> $TP := \left(1 + (Z-1)\cdot B^2\cdot f1^2\right)\cdot T + R\cdot(Z-1)\cdot f1\cdot f2\cdot B^2\cdot TT + \left(-\frac{B^2\cdot Z\cdot f1}{c1}\right)\cdot X;$

$$TP := \left(1 + \frac{\left(\frac{1}{\sqrt{-\frac{1}{\frac{c1^2}{v^2} + \frac{c2^2}{k^2\,v^2}} + 1}} - 1\right) c1^2}{\left(\frac{c1^2}{v^2} + \frac{c2^2}{k^2\,v^2}\right) v^2}\right) T + \frac{c2^2 \left(\frac{1}{\sqrt{-\frac{1}{\frac{c1^2}{v^2} + \frac{c2^2}{k^2\,v^2}} + 1}} - 1\right) TT}{v^2\,k \left(\frac{c1^2}{v^2} + \frac{c2^2}{k^2\,v^2}\right)} \quad (6)$$

> $TTP := \frac{(Z-1)}{R} \cdot f1 \cdot f2 \cdot B^2 \cdot T + \left(1 + (Z-1) \cdot f2^2 \cdot B^2\right) \cdot TT + \left(-\frac{B^2 \cdot Z \cdot f2}{c2}\right) \cdot X;$

$$TTP := \frac{\left(\frac{1}{\sqrt{-\frac{1}{\frac{c1^2}{v^2}+\frac{c2^2}{k^2 v^2}}+1}}-1\right) c1^2\, T}{v^2 k \left(\frac{c1^2}{v^2}+\frac{c2^2}{k^2 v^2}\right)} + \left(1 + \frac{\left(\frac{1}{\sqrt{-\frac{1}{\frac{c1^2}{v^2}+\frac{c2^2}{k^2 v^2}}+1}}-1\right) c2^2}{k^2 v^2 \left(\frac{c1^2}{v^2}+\frac{c2^2}{k^2 v^2}\right)}\right) TT - \frac{X}{\left(\frac{c1^2}{v^2}+\frac{c2^2}{k^2 v^2}\right)\sqrt{-\frac{1}{\frac{c1^2}{v^2}+\frac{c2^2}{k^2 v^2}}+1}\; k\, v} \quad (7)$$

> $XP := \left(-B^2 \cdot c1 \cdot f1 \cdot Z\right) \cdot T + \left(-B^2 \cdot f2 \cdot c2 \cdot Z\right) \cdot TT + Z \cdot X;$

$$XP := -\frac{T\, c1^2}{\left(\frac{c1^2}{v^2}+\frac{c2^2}{k^2 v^2}\right)\sqrt{-\frac{1}{\frac{c1^2}{v^2}+\frac{c2^2}{k^2 v^2}}+1}\; v} - \frac{TT\, c2^2}{\left(\frac{c1^2}{v^2}+\frac{c2^2}{k^2 v^2}\right)\sqrt{-\frac{1}{\frac{c1^2}{v^2}+\frac{c2^2}{k^2 v^2}}+1}\; k\, v} + \frac{X}{\sqrt{-\frac{1}{\frac{c1^2}{v^2}+\frac{c2^2}{k^2 v^2}}+1}} \quad (8)$$

> $w := k \cdot v;$

$$w := k\, v \quad (9)$$

> $nTP := simplify(TP);$

**(10)**

$$nTP := \frac{1}{\sqrt{\frac{c1^2 k^2 - k^2 v^2 + c2^2}{c1^2 k^2 + c2^2}} \left(c1^2 k^2 + c2^2\right)} \left( -c2^2 k\, TT \sqrt{\frac{c1^2 k^2 - k^2 v^2 + c2^2}{c1^2 k^2 + c2^2}} + T\, c1^2 k^2 + \sqrt{\frac{c1^2 k^2 - k^2 v^2 + c2^2}{c1^2 k^2 + c2^2}}\, T\, c2^2 + c2^2 k\, TT - k^2 v X \right) \qquad (10)$$

> $nTTP := simplify(TTP);$

$$nTTP := -\frac{k\, c1^2 \left(-TT\, k + T\right) \sqrt{\frac{c1^2 k^2 - k^2 v^2 + c2^2}{c1^2 k^2 + c2^2}} + \left(-T\, c1^2 + X v\right) k - TT\, c2^2}{\sqrt{\frac{c1^2 k^2 - k^2 v^2 + c2^2}{c1^2 k^2 + c2^2}} \left(c1^2 k^2 + c2^2\right)} \qquad (11)$$

> $nXP := simplify(XP);$

$$nXP := \frac{-c1^2 \left(T v - X\right) k^2 - k v\, TT\, c2^2 + X\, c2^2}{\sqrt{\frac{c1^2 k^2 - k^2 v^2 + c2^2}{c1^2 k^2 + c2^2}} \left(c1^2 k^2 + c2^2\right)} \qquad (12)$$

>

Let us manually simplify the Maplesoft™ output further. Before we can do that, we write down the mapping of Maplesoft variables with respect to the actual variables used in the text.

$T = t_1$ , $TT = t_2$ , $X = x$ , $nTP = t'_1$ , $nTTP = t'_2$ , $nXP = x'$

We use $t_1 = k\, t_2$

**<u>Transformation of $t_1$</u>**

$$nTP := \frac{1}{\sqrt{\frac{c1^2 k^2 - k^2 v^2 + c2^2}{c1^2 k^2 + c2^2}} \left(c1^2 k^2 + c2^2\right)} \left( -c2^2 k\, TT \sqrt{\frac{c1^2 k^2 - k^2 v^2 + c2^2}{c1^2 k^2 + c2^2}} + T\, c1^2 k^2 + \sqrt{\frac{c1^2 k^2 - k^2 v^2 + c2^2}{c1^2 k^2 + c2^2}}\, T\, c2^2 + c2^2 k\, TT - k^2 v X \right) \qquad (10)$$

With $t_1 = k\, t_2$, the two square root terms in the parenthesis in the numerator, cancel each other and the expression simplifies to

$$t_1' = \frac{t_1 c_1^2 k^2 + t_2 c_2^2 k - k^2 v x}{(c_1^2 k^2 + c_2^2)\sqrt{1 - \dfrac{k^2 v^2}{(c_1^2 k^2 + c_2^2)}}}$$

$$= \frac{t_1 (c_1^2 k^2 + c_2^2) - k^2 v x}{(c_1^2 k^2 + c_2^2)\sqrt{1 - \dfrac{k^2 v^2}{(c_1^2 k^2 + c_2^2)}}}$$

$$= \frac{t_1 - \dfrac{k^2 v x}{(c_1^2 k^2 + c_2^2)}}{\sqrt{1 - \dfrac{k^2 v^2}{(c_1^2 k^2 + c_2^2)}}}$$

$$= \frac{t_1 - \dfrac{v x}{(c_1^2 + c_2^2 / k^2)}}{\sqrt{1 - \dfrac{v^2}{(c_1^2 + c_2^2 / k^2)}}}$$

$$= \frac{t_1 - \dfrac{v x}{c_e^2}}{\sqrt{1 - \dfrac{v^2}{c_e^2}}} \quad \text{where } c_e = \sqrt{c_1^2 + c_2^2 / k^2}$$

**Transformation of $t_2$**

$$nTTP := -\frac{k\, c1^2 (-TT\, k + T)\sqrt{\dfrac{c1^2 k^2 - k^2 v^2 + c2^2}{c1^2 k^2 + c2^2}} + (-T\, c1^2 + X v)\, k - TT\, c2^2}{\sqrt{\dfrac{c1^2 k^2 - k^2 v^2 + c2^2}{c1^2 k^2 + c2^2}}\,(c1^2 k^2 + c2^2)} \tag{11}$$

With $t_1 = k\, t_2$, the first term in the numerator drops. We get,

$$t_2' = \frac{t_2 (c_1^2 k^2 + c_2^2) - x w}{\sqrt{1 - \dfrac{k^2 v^2}{c_1^2 k^2 + c_2^2}}\,(c_1^2 k^2 + c_2^2)}$$

$$t_2' = \frac{t_2 - \dfrac{x w}{(c_1^2 k^2 + c_2^2)}}{\sqrt{1 - \dfrac{k^2 v^2}{c_1^2 k^2 + c_2^2}}} = \frac{t_2 - \dfrac{x w}{k^2 c_e^2}}{\sqrt{1 - \dfrac{w^2}{k^2 c_e^2}}} = \frac{t_2 - \dfrac{x w}{k^2 c_e^2}}{\sqrt{1 - \dfrac{v^2}{c_e^2}}} = t_1' / k \text{ where, } c_e^2 = (c_1^2 + c_2^2 / k^2)$$

**Transformation of *x***

$$nXP := \frac{-c1^2\,(T\,v - X)\,k^2 - k\,v\,TT\,c2^2 + X\,c2^2}{\sqrt{\frac{c1^2\,k^2 - k^2\,v^2 + c2^2}{c1^2\,k^2 + c2^2}}\,\left(c1^2\,k^2 + c2^2\right)} \quad \textbf{(12)}$$

With $t_1 = k\,t_2$ we get,

$$x' = \frac{x(c_1{}^2k^2 + c_2{}^2) - vt(c_1{}^2k^2 + c_2{}^2)}{\sqrt{1 - \frac{k^2v^2}{c_1{}^2k^2 + c_2{}^2}}\,(c_1{}^2k^2 + c_2{}^2)}$$

$$= \frac{x - vt}{\sqrt{1 - \frac{v^2}{c_e{}^2}}} \text{ where } c_e = \sqrt{c_1{}^2 + c_2{}^2 / k^2}$$

**Appendix C**

Showing $s'^2 = c_1^{\,2}t_1'^{\,2} + c_2^{\,2}t_2'^{\,2} - x'^2 - y'^2 - z'^2 = c_1^{\,2}t_1^{\,2} + c_2^{\,2}t_2^{\,2} - x^2 - y^2 - z^2 = s^2$

using simplified expressions of eq. (13) and Maplesoft™ [21].

> $t1 := t2 \cdot k;$

$$t1 := t2\,k \quad \textbf{(1)}$$

> $ce := \mathrm{sqrt}\left(c1^2 + \frac{c2^2}{k^2}\right);$

$$ce := \sqrt{c1^2 + \frac{c2^2}{k^2}} \quad \textbf{(2)}$$

> $t1p := \frac{\left(t1 - \frac{x \cdot v}{ce^2}\right)}{\mathrm{sqrt}\left(1 - \frac{v^2}{ce^2}\right)};$

$$t1p := \frac{t2\,k - \frac{x\,v}{c1^2 + \frac{c2^2}{k^2}}}{\sqrt{1 - \frac{v^2}{c1^2 + \frac{c2^2}{k^2}}}} \quad \textbf{(3)}$$

> $t2p := \frac{\left(t2 - \frac{x \cdot v}{k \cdot ce^2}\right)}{\mathrm{sqrt}\left(1 - \frac{v^2}{ce^2}\right)};$

$$t2p := \frac{t2 - \frac{x\,v}{k\left(c1^2 + \frac{c2^2}{k^2}\right)}}{\sqrt{1 - \frac{v^2}{c1^2 + \frac{c2^2}{k^2}}}} \quad \textbf{(4)}$$

> $xp := \frac{(x - v \cdot t1)}{\mathrm{sqrt}\left(1 - \frac{v^2}{ce^2}\right)};$

$$xp := \frac{-k\,t2\,v + x}{\sqrt{1 - \frac{v^2}{c1^2 + \frac{c2^2}{k^2}}}} \tag{5}$$

>

> $dsp2 := c1^2 \cdot t1p^2 + c2^2 \cdot t2p^2 - xp^2;$

$$dsp2 := \frac{c1^2 \left(t2\,k - \frac{x\,v}{c1^2 + \frac{c2^2}{k^2}}\right)^2}{1 - \frac{v^2}{c1^2 + \frac{c2^2}{k^2}}} + \frac{c2^2 \left(t2 - \frac{x\,v}{k\left(c1^2 + \frac{c2^2}{k^2}\right)}\right)^2}{1 - \frac{v^2}{c1^2 + \frac{c2^2}{k^2}}} - \frac{(-k\,t2\,v + x)^2}{1 - \frac{v^2}{c1^2 + \frac{c2^2}{k^2}}} \tag{6}$$

> $simplify(dsp2);$

$$\left(c1^2\,k^2 + c2^2\right) t2^2 - x^2 \tag{7}$$

>

In the above part generated by Maplesoft™, we used symbols that map to variable symbols used in the text as follows,

$$c1 = c_1, c2 = c_2, t1 = t_1, t2 = t_2$$
$$t1p = t_1', t2p = t_2', x1p = x'$$

Using $k.t_2 = t_1$ in (7) we get

$$s'^2 = c_1'^2 t_1'^2 + c_2'^2 t_2'^2 - x'^2 = c_1^{\,2} t_1^2 + c_2^{\,2} t_2^2 - x^2 = s^2$$

**Appendix D**

Consider (eq. (38))

$$(\frac{(E_2)_1}{c_1})^2 + (\frac{(E_2)_2}{c_2})^2 - m_0{}^2 \bar{u}_2{}^2 = m_0^2 c_2^2 \qquad \text{(D.1)}$$

From the eq.(39) box , we have

$$\frac{(E_1)_1}{(E_2)_1} = \frac{(E_1)_2}{(E_2)_2} = \frac{1}{\rho}\ ,\ \bar{u}_2 = \gamma_2 \bar{v}_2 = \gamma_2 k \bar{v}_1 = \rho\gamma_1 \bar{v}_1 \text{ and } \rho = \frac{c_2}{c_1}$$

or,

$$(E_2)_1 = \rho(E_1)_1$$

$$(E_2)_2 = \rho(E_1)_2$$

So (D.1) becomes,

$$(\frac{\rho(E_1)_1}{c_1})^2 + (\frac{\rho(E_1)_2}{c_2})^2 - m_0{}^2 (\rho\gamma_1 \bar{v}_1)^2 = m_0^2 c_2^2$$

or,

$$(\frac{(E_1)_1}{c_1})^2 + (\frac{(E_1)_2}{c_2})^2 - m_0{}^2 (\gamma_1 \bar{v}_1)^2 = m_0^2 c_2^2 / \rho^2$$

or,

$$(\frac{(E_1)_1}{c_1})^2 + (\frac{(E_1)_2}{c_2})^2 - m_0{}^2 \bar{u}_1{}^2 = m_0^2 c_1^2$$

**Brief Biography and Declaration**

Dr. Sajjad Zahir is a Professor Emeritus, University of Lethbridge, Lethbridge, Alberta, Canada.
He completed graduate education both in Theoretical Physics (Ph. D.) and Computer Science (MS) from the University of Oregon, Eugene, USA and previously held research positions in Physics in several Canadian Universities. Then he worked in the areas of Decision Sciences and Information Systems areas at the University of Lethbridge (UL) as a tenured Assistant, Associate, and Full Professor and took an early retirement. He has published many research articles in physics, operation research, decision sciences, and information systems. He is still affiliated with the University of Lethbridge as Professor Emeritus and, in his own interest, active again in physics research. However, he is solely responsible for the content of his current physics research, and thus results of his current physics research should have no bearing on the Physics Department of the University of Lethbridge where he never worked. Dr. Zahir fully understands and respects the reputational achievement of the University of Lethbridge and would remain careful not to impact its image. For full information about Dr. Sajjad Zahir's background, education, employment history, and research records see his personal website https://sites.google.com/view/sajjadzahir/SajjadZahir.